%% file: GRTC.tex
\documentclass[fleqn,11pt]{article}

\usepackage{amsfonts,amsmath,amssymb}
\usepackage{latexsym}

\input preamble.tex

\newcommand{\veps}{\varepsilon}

\begin{document}
\twocolumn[ \jnumber{31 (3)}{2025}

\Title{Gauge gravitation theory in Riemann--Cartan space-time\yy and the nonsingular
Universe}

\Aunames{A.V. Minkevich\au{a,b,1}}

\Addresses{ \adr a {Department of Theoretical Physics and Astrophysics, Belarussian State
University, Minsk, Belarus} \adr b {Department of Mathematics and Informatics, University
of Warmia and Mazury in Olsztyn, Poland}}

\Abstract
   {The gauge gravitation theory in the Riemann--Cartan space-time is investigated in order to
solve the fundamental problems of the general relativity theory. The constraints for indefinite
parameters of the theory under which solutions of isotropic cosmology describe a
nonsingular accelerating Universe are given. Numerical solutions of cosmological equations
near the limiting energy density by transition from gravitational compression to expansion in
dependence on energy density in the case of flat, closed and open models are obtained.
Some physical consequences of gauge gravitational theory in the Riemann--Cartan
space-time in astrophysics are discussed.}
\medskip

] 
\email 1 {minkav@tut.by\\ \cm (Corresponding author)}

\section{Introduction}

The gauge gravitation theory in the Riemann--Cartan space-time (GTRC) is a direct
development of the general relativity theory (GR), in which Einstein's idea of the dependence
of the properties of physical space-time from the distribution and motion of matter finds its
further development: in addition to curvature, space-time has torsion. The GTRC was
created as Poincar\'{e} gauge theory of gravity (PGTG) and it has quite a long history
starting with the classic works of T.W.B. Kibble, D.W. Sciama, D. Ivanenko, A. Trautman
and others (see, e.g. \cite{a1,a2,a3,a4,a5,a6} and references therein). Usually, the
Poincar\'{e} group as a group of coordinate transformations in the form of a semidirect
product of a translations group and a group of Lorentz coordinate transformations in
Minkowski space-time is considered as the gauge group of PGTG. At the same time, the
equations of GTRC can be obtained by considering the direct product of the 4-parametric
group of space-time translations  and the 6-parametric group of tetrad Lorentz
transformations as a gauge group. \footnote{Although GTRC is often referred in the
literature as PGTG, there is an important difference between the two. Note that the
4-parametric group of space-time translations contains arbitrary space-time transformations,
including the Poincar\'{e} transformations.} As a result, the orthonormal tetrad and
nonholonomic Lorentz connection play the role of gravitational field variables in GTRC and
torsion and curvature tensors play a role as gravitational field strengths. While the
gravitational Lagrangian is an invariant function constructed using tensors of gravitational
field strength, the Lagrangian of matter interacting with the gravitational field is constructed
on the basis of the corresponding special-relativistic Lagrangian by replacing partial
derivatives of material variables in Galilean coordinates in Minkowski space-time with
covariant derivatives. Depending on the use of the covariant derivative, determined using full
nonholonomic connectivity or Riemannian (Christoffel brackets) connectivity, the interaction
is minimal or non-minimal.\footnote{The covariant derivative can be constructed using
nonholonomic Lorentz connection and Christoffel brackets, which leads to a different
non-minimal coupling of the gravitational field with matter, which is not considered further.}
The Noether invariants corresponding to two symmetry subgroups of GTRC, namely the
canonical energy-momentum tensor and the tensor of tetrad spin moment play the role of
sources of the gravitational field in the frame of GTRC. Their magnitude depends on which
connection of the gravitational field with matter---minimal or non-minimal---is used. In
particular, the tetrad spin moment characterizes the spin properties of matter only in the case
of minimal coupling of the gravitational field with matter. In the case of the Riemannian
coupling used in GR, the tetrad spin moment disappears (equal to zero). When using minimal
coupling with the gravitational field, spinor fields make an important contribution to the tetrad
spin moment.\footnote{The interaction of electromagnetic and Yang-Mills fields with the
gravitational field in the framework of GTRC is set using Riemannian connectivity in order to
preserve gauge invariance for interaction of these fields.} As the consistent variational
formalism for so-called spinning matter in space-time with curvature and torsion shows, the
rotational moment of spinning matter by using minimal connection with gravitational field
manifests itself as a spin moment equal to the tetrad spin moment in the framework of GTRC
\cite{a7,a8}. It is important because various star systems, galaxies, and clusters of galaxies
have large rotational moments. Currently, GTRC is one of the most important directions in
the development of the gravitation theory, which opens up the possibility of solving
fundamental problems of GR.

In spite of a large number of cosmological investigations during the last time the Big Bang
cosmological scenario of GR remains the principal model of modern cosmology. The
principal problem of Big Bang scenario is cosmological problem---the problem of the
beginning of the Universe in time connected directly with cosmological singularity,
gravitational singularity with divergent energy density. After the proof of the
Penrose-Hawking theorems on the inevitability of singularities in GR, the attitude towards
gravitational singularities among many researchers acquired a character as an inevitable
reality. At the same time, many attempts have been made to solve the problem of gravitational
singularities. Solving the problem of cosmological singularity does not just mean obtaining
particular regular solutions, but excluding possible singular solutions from physical grounds.
The GTRC opens up possibilities for solving this cosmological problem by classical
description of gravitational field due to the conclusion about the possible existence of a
limiting (the maximum allowed) energy density in the Nature, near which the gravitational
interaction has the character of repulsion (see \cite{k1} and references therein). As result
gravitational singularity with divergent energy density is impossible and all cosmological
models of isotropic cosmology filled by usual gravitating matter by certain restrictions on
indefinite parameters are regular. The stage of cosmological contraction was preceded the
stage of cosmological expansion and there are no restrictions on the existence of the
Universe in time, both in the past and in the future. The physical processes occurring in
matter at the beginning of cosmological expansion of the hot Universe described in
accordance with the theory of elementary particles are depending essentially on the limiting
energy density (limiting temperature), the value of which should exceed the energy density in
the densest astrophysical objects and be less than the Planck energy density. The absence of
the beginning of the Universe in time can lead to changes in the history of the early Universe
associated with the removal of restrictions inherent in the standard cosmological Big Bang
scenario.  In this regard, we would like to point out the possible corrections in the history of
the early Universe revealed in observations of the James Webb Space Telescope, in
particular, the observation puzzles of massive bright galaxies in the early Universe. The
GTRC leads also to the gravitational repulsion at cosmological asymptotics, when energy
density in the Universe is very small in comparison with limiting energy density, and explains
the cosmological acceleration at modern epoch as vacuum effect without using any dark
energy, although the cosmological equations in this case have the form of Friedmann
cosmological equations with an effective cosmological constant \cite{k2}. All this points to
the need to study isotropic cosmology in the frame of GTRC, physical processes at different
stages of the evolution of the Universe, their dependence on indefinite parameters in
comparison with standard $\Lambda$CDM-model. Some such restrictions were used in our
previous papers. Further research requires clarification of these limitations.

This article is devoted to the analysis of solutions of isotropic cosmology in dependence of
indefinite parameters. Initially, cosmological solutions are considered when the energy density
is small compared to the limiting energy density, where the vacuum effect of gravitational
repulsion can play an important role. A comparative analysis of cosmological solutions in the
field of limiting energy density for flat, closed and open models is carried out. Further, some
physical consequences of GTRC in astrophysics are discussed.

\section{Equations of isotropic cosmology in Riemann--Cartan space-time and asymptotics of cosmological models}

Equations of isotropic cosmology were obtained in the frame of GTRC based on general
expression of gravitational Lagrangian contained both the scalar curvature and various
invariants quadratic  in the curvature and torsion tensors with indefinite parameters by
assumption of parity conservation (so without using Levi-Civita symbol). The system of
gravitational equations is a complex system of differential equations in partial derivatives that
makes it possible to find the gravitational field (tetrad or metrics and nonholonomic Lorentz
connection) generated by given material systems with a certain distribution of energy,
momentum, spin  moments \cite{a5, a6, a8}. Note that the torsion of space-time can be
generated not only by the spin of elementary particles (see \cite{a9}), which has a quantum
nature, but also by the classical spin moment in the form of rotation moment introduced
within the framework of the Noether formalism of classical field theory, as well as by the
energy-momentum tensor. Moreover, the physical vacuum has torsion under certain
constraints on the indefinite parameters of the gravitational Lagrangian \cite{k2}. Because the
average value of the spin moment in the frame of isotropic cosmology is equal to zero, the
space-time torsion is created only by energy-momentum tensor, it means by energy density
$\veps$ and pressure $p$ of gravitating matter. Equations of isotropic cosmology take the
following form \cite{k1, k2, k3,k4,k5}\footnote{The most part of definitions and notations
of our previous work (see, e.g., \cite{k1}) are used below besides: the scale factor of
Robertson-Walker metric is denoted by $a(t)$ instead $R(t)$, it will be written $\alpha_G$
instead the parameter $\alpha$.}
\begin{eqnarray}\label{2.1}
\frac{k c^2}{a^2} + (H-2S_1)^2 -S_2^2 \nonumber\\
=\frac{1}{{6f_0 Z}}
        \left[
            {\veps  -6 b S_2^2
            + \frac{\alpha_G }{4} \left( {\veps  - 3p - 12bS_2^2 } \right)^2 }
        \right],
\end{eqnarray}
\begin{eqnarray}\label{2.2}
    \dot{H}-2\dot{S}_1 +H (H-2S_1) \nonumber\\
    =-\frac{1} {{12f_0 Z}}
        \left[
            \veps  + 3p - \frac{\alpha_G } {2} \left( {\veps - 3p - 12bS_2^2 } \right)^2
        \right],
\end{eqnarray}
where $H=\dot{a}/a $  (a dot denotes the differentiation with respect to time $ t$),
$k=+1,0,-1$ for closed, flat and open models respectively and $Z=1+\alpha_G\left( \veps -
3p - 12b S_2^2\right)$, the torsion functions $S_1$ and $S_2$ are:
\begin{eqnarray}\label{2.3}
    S_1  = -\frac{\alpha_G }{4Z} [\dot \veps
    - 3 \dot p + 12f_0 \omega H S_2^2 \nonumber\\
    -12( {2b - \omega f_0 } ) S_2 \dot S_2],
\end{eqnarray}
\begin{eqnarray}\label{2.4}
 S_{2}^{2}  = \frac{\veps - 3p}{12b} + \frac
{1-(b/2f_0) (1 +  \sqrt{X})} {12b \alpha_G (1- \omega/4)},
\end{eqnarray}
where $X=1+ \omega (f_0^2/b^2) [1- (b/f_0) - 2(1- \omega /4) \alpha_G ( \veps+ 3p)]\ge 0$
and $f_0=\frac{c^{2}}{16\pi G}$ ($G$ is Newton's gravitational constant), $\alpha_G,
\omega, b$ are indefinite parameters, wherein $\alpha_G$ has inverse dimension of energy
density, $\omega$ without dimension and dimension of $b$ is the same as $f_0$. The
parameters  $\alpha_G, \omega$ are connected with quadratic in the curvature terms of
gravitational Lagrangian and the parameter $b$ with terms quadratic in the torsion tensor.

The analysis of homogeneous isotropic cosmological models (HICM) is based on
cosmological equations (1)--(2) by using the equation of the energy conservation law, which
has the same form as in GR
\begin{equation}
\dot{\veps}+3H\left(\veps+ p\right)=0.
\end{equation}
Equation (5) is fulfilled in the framework of the GTRC due to the fact that, in accordance
with the equations of dynamics for spinning matter in space-time with curvature and torsion
when using a minimal coupling of the gravitational field with matter, the Riemannian
divergence of the metric energy-momentum tensor is zero \cite{a7}. The dynamics of HICM
depends on energy density regardless of her origin and equation of state ${p}= {p
({\veps})}$; in the case of the presence of scalar fields (inflationary models) knowledge of
their potential is necessary  \cite{k4}. The behavior of cosmological solutions depends on
restrictions on parameters $\alpha_G, \omega, b$ and their most important properties arise if
the value of parameter ${\alpha_G}^{-1}$ corresponds to some high energy density, by
which at cosmological asymptotics $\alpha_G \veps \ll 1$ and $0<1-\frac{b}{f_0}\ll 1$.
Under such assumptions, the HICM were studied in \cite{k1}, where the condition $0
<\omega\ll 1$ was additionally assumed for simplicity. As was shown, cosmological
solutions in the case of flat models describe regular accelerating Universe with limiting energy
density, the presence of which is secured by the condition $X\ge 0$ at  $0 <\omega<4$.
However, as will be shown below, similar situation takes place, if the parameter $\omega$ is
not small and $\omega\sim 1$. Indeed, we have at asymptotics according to (4) in the first
approximation with respect to small parameter $ x=1-b/f_0 $ regardless of the acceptable
values of $\omega$ :
\begin{equation}
 S_{2}^{2}  = \frac{\veps - 3p}{12b} + \frac{1-(b/f_0)} {12b \alpha_G }.
\end{equation}
As a result, cosmological equations (1)--(2) at asymptotics take the form of Friedmann
cosmological equations with an effective cosmological constant $\Lambda=\frac{(1 -
\frac{b}{f_0})^2} {8b\alpha_G}$ induced by the torsion function (6):
\begin{equation}
    \frac{k c^2}{a^2 } + H^2  = \frac{1}{6b }\left[\veps + \frac{1}{4\alpha_G} \left(1 - \frac{b}{f_0}\right)^2
     \right],
\end{equation}
\begin{equation}
    \dot H + H^2  =  - \frac{1} {{12b }}\left[ (\veps + 3p) - \frac{1}{2\alpha_G}
    \left(1 - \frac{b}{f_0}\right)^2 \right].
\end{equation}
It follows from (7)--(8) that the vacuum energy density is ${\veps_{\rm vac}}=
\frac{1}{4\alpha_G} \left(1 - \frac{b}{f_0}\right)^2 $. This is true not only in the case of flat
cosmological models studied in \cite{k1}, but also in the case of models with non-Euclidean
topology ($k=+1,-1$), if the gravitating vacuum is determined on the basis of HICM when
the energy density of matter tends to zero \cite{k2}. If $\veps_{\rm vac}\ll \veps\ll
\veps_{\rm max}$, where $\veps_{\rm max}$ is the limiting energy density, the dynamics of
HICM described by cosmological equations (1)--(2) coincides practically with that of
Friedmannian cosmology since the value of the parameter $b$ is close to $f_0$. We see that
the parameter $\omega$, which is important at the beginning of the cosmological expansion
near the limiting energy density, does not affect the evolution of HICM at asymptotics
regardless of its acceptable values ($0<\omega<4$). As the energy density decreases ($
\veps \sim\veps_{\rm vac}$), the vacuum effect of gravitational repulsion becomes
important,  the sign of the cosmological acceleration changes according to equation (8), and
with a further decrease in the energy density, the gravitational interaction has the character of
repulsion, depending on the magnitude $\veps$. Further, the dynamics of HICM-models
under extreme conditions near the limiting energy density will be investigated.

\section{Dynamics of homogeneous isotropic cosmological models near limiting energy density}

Solutions of the cosmological equations (1)--(2) in an analytical form were obtained in
\cite{k1}, where their numerical analysis was carried out near the limiting energy density for
flat cosmological models under the condition $0 <\omega\ll 1$. To carry out a general
analysis of solutions obtained without using these restrictions, we will first write them down
like \cite{k1} in dimensionless form, analyze when $\omega =1$ and then consider their
dependence on parameter $\omega$. The dimensionless form of solutions is obtained by
using transitions to dimensionless values noted by means of tilde:
\[
    t \to\tilde{t}=t /\sqrt{6f_0 \omega \alpha_G},
\]
\[
    H\to\tilde{H}=H\sqrt{6f_0 \omega \alpha_G},
\]
\[
 \veps\to\tilde{\veps}=\omega \alpha_G\, \veps,
\]
\[
 p\to\tilde{p}=\omega \alpha_G\,p,
\]
\[
 S_{1,2}\to\tilde{S}_{1,2}=S_{1,2}\sqrt{6f_0 \omega \alpha_G},
\]
\[
 b\to\tilde{b} = b/f_0,
\]
\[
 a\to\tilde{a}=a/{c \sqrt{6f_0 \omega \alpha_G}},
\]
\[
 \tilde{\veps}'+3\tilde{H}\left(\tilde{\veps}+\tilde{p}\right)=0,
\]
where prim denotes the differentiation with respect to $\tilde{t}$. Then the Hubble parameter
can be written as:
\begin{equation}
\tilde {H}=\tilde{H}_{\pm}= \pm \frac{ \sqrt{\tilde{A_1}}}{1+\frac{3}{2\tilde{b}Z}\tilde{D}},
\end{equation}
where
\begin{eqnarray}
\tilde {A_1}=\frac{\tilde {\veps}- 3\tilde {p}}{2\tilde {b}} + \frac{\tilde {\veps}+ 3\tilde {p}}{2 Z}
\nonumber\\
{}+\omega \frac{1-(\tilde {b}/2)(1+ \sqrt{X})}{2\tilde {b} (1-\omega/4)}\left(1-\frac{\tilde
{b}}{Z}\right)
\nonumber\\
{}+\omega \frac{(1-(\tilde {b}/2)(1+ \sqrt{X}))^2}{4{Z} (1-\omega/4)^2} -  \frac{k}{\tilde {a}^2},
\end{eqnarray}

\begin{eqnarray}\label{36}
\tilde {D} = \frac{1}{2} \left(3\frac{d \tilde p}{d \tilde\veps}-1 \right) \left( \tilde {\veps}+ \tilde {p} \right)
 +\frac{1}{3}\left(\tilde {\veps} - 3\tilde { p} \right)
\nonumber\\
{}-\frac{\omega \tilde {b}}{6 (1-\omega/4)} \sqrt{X}
+ \frac{\omega ( 1-\tilde { b }/2 )}{3 (1-\omega/4)}
\nonumber\\
{}+\frac{1-\omega (1/2 \tilde {b})}{2\sqrt{X}}\left[\left(3\frac{d \tilde {p}}{d \tilde {\veps }}+1
\right) \left(\tilde { \veps }+ \tilde {p }\right) \right].
\end{eqnarray}

The quantities $X$ and $Z$ are dimensionless and can be written in the form:
\begin{eqnarray}
X=1+ \frac {\omega} {\tilde {b}^2} (1 - \tilde {b}) - 4\frac{1- \omega /4}{\tilde {b}^2}
{\tilde {\veps}},  \nonumber\\
Z =  \frac{-\omega/4 + (\tilde {b}/2)(1+ \sqrt{X})} {1-\omega/4}.
\end{eqnarray}

The time derivative of the Hubble parameter is:
\begin{eqnarray}
 \tilde {H}'= - \tilde {H}^2+\Biggl[ \tilde {A}_2+\frac{9}{2 \tilde{b}Z} ( \tilde {\veps}+ \tilde {p})  \tilde {H}^2 \Biggl(D_1
\nonumber\\
{}+\frac{ \tilde {D}}{2 \tilde {b}Z \sqrt{X}}\left (1+3\frac{d \tilde{ p} }{d \tilde { \veps} }\right ) \Biggr) \Biggr]
\left [1+\frac{3 \tilde{D} }{2 \tilde {b}Z}\right]^{-1},
\end{eqnarray}
where
\begin{equation}
 \tilde {A}_2=-\frac{1}{2Z}\left [ \tilde {\veps}+3 \tilde {p}- \omega \frac{(1-( \tilde {b}/2) (1+ \sqrt{X}))^2}{2  (1-\omega/4)^2} \right],
\end{equation}
\begin{eqnarray}
D_1= \frac{1}{6} \left(3\frac{d \tilde p}{d \tilde\veps}-1 \right) \left(3\frac{d \tilde p}{d \tilde\veps}+1 \right)
\nonumber\\
{}+ \frac{3}{2} \left(\tilde\veps+\tilde p \right)
\frac{d^2\tilde p}{d^2\tilde \veps}
+\frac{\omega}{6\tilde b \sqrt{X}}\left(1+3\frac{d \tilde p}{d \tilde\veps}\right)
\nonumber\\
 {}+ \frac{1-\frac{\omega}{2\tilde b}}{2\sqrt{X}}\left[\left(1+\frac{d \tilde p}{d \tilde\veps}\right)
 \left(1+3\frac{d \tilde p}{d \tilde \veps}\right) + 3(\tilde\veps+\tilde p)\frac{d^2\tilde p}{d^2\tilde\veps}\right]
\nonumber\\
{}+ \frac{1-\frac{\omega}{2\tilde b}}{2X^{3/2}}\frac{1}{\tilde b^2} (\tilde\veps+\tilde p)
\left(1+3\frac{d \tilde p}{d \tilde\veps}\right)^2 (1-\omega/4).
\end{eqnarray}

Now by using equations (9)--(15) we will investigate HICM at the beginning of cosmological
expansion of the hot Universe near limiting energy density in the case $\omega =1$. We will
use the equation of state for ultrarelativistic matter $\tilde {p}=\tilde {\veps}/3$, then
according to equation (5) we have ${\veps} a^4={\rm const}$ and as result
$k/\tilde{a}^2=k\sqrt{\tilde{\veps}}/ C_1$, where $C_1$ is the constant,  the limitations for
which in the case of a closed cosmological models follow from the requirement $A_1\ge 0$.
The difference between the parameters $b$ and $f_0$ is small by virtue the condition used
$0<1-\frac{b}{f_0}\ll 1$ and it is significant only at cosmological asymptotics, where an
effective cosmological constant can play an important role. In this regard, when performing
numerical calculations near limiting energy density, we will think that $\tilde{b}=1$ and then
$\tilde{\veps}_{\rm max}=1/4(1- {\omega}/4)= \frac{1}{3}$. The results of numerical
analysis in the case of flat models ($k=0$) and closed models ($k=+1$, $C_1=15$) are
presented in Fig.~1--Fig.~3 with an accuracy of 0.001. These results are close to them
obtained for flat models in  \cite{k1}. Strictly speaking, Fig.~1 and Fig.~2 depict the
behavior of the corresponding quantities only in the region of extremely high energy densities
near $\tilde{\veps}_{\rm max}$ and they are not applicable in asymptotics due to the
condition used $\tilde{b}=1$, which excludes the appearance of an effective cosmological
constant, as well as due to a change in the equation of state of matter with a change in energy
density. However, due to the weak dependence of the results on the equation of state $p = w
\veps$ ($0 \le w \le 1/3$), these figures reflect some features of cosmological models in
asymptotics. So, in the case of flat models ($k=0$), the state with coordinates $(0, 0)$ in
Fig.~1 is achieved at $t \to \pm \infty$, while in the case of closed models ($k=+1$) this state
is not achievable. In fact, in the case of flat models, the Hubble parameter tends at $t \to \pm
\infty$ to its vacuum value $\tilde{H}^{\rm (vac)}_\pm= \pm (1/2) (1- \tilde{b})$, and in the
case of closed models the transition from compression to expansion at some energy density
$\tilde{\veps_c}$ takes place. The value $\tilde{\veps_c}$ depends on an unknown value
$C_1$ and it decreases (increases) with the increase (decrease) of the value of $C_1$. Since
the value of the energy density $\tilde{\veps}$ at asymptotics is many orders of magnitude
less than $\tilde{\veps}_{\rm max}$, it is strictly speaking not possible to depict the behavior
of the quantities in question on graphs at asymptotics and the value used $C_1 = 15$ is taken
to clarify the interpretation of behavior of the Hubble parameter in the case of a closed model
at small values $\tilde{\veps}$.

  The numerical data obtained for flat cosmological models are given below. It follows from Fig.~1 that the parameter $\tilde{H}_{+}$ ($\tilde{H}_{-}$)
 vanishing at the limiting energy density $\tilde{\veps}_{\rm max}$ reaches its maximum (minimum)
 value $\tilde{H}_{+}= 0.314$ ($\tilde{H}_{-}= - 0.314$) at $\tilde{\veps}_1=0.185= 0.556\, \tilde{\veps}_{\rm max}$.

\begin{figure}
\centering
\includegraphics[width=\columnwidth]{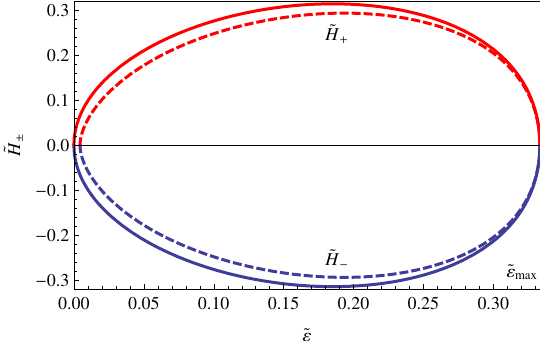}
\caption{Parameter $\tilde{H}=\tilde{H}_\pm$ as function of $\tilde{\veps}$ for flat model (solid lines) and for closed model (dashed lines): $\tilde{H}_+$ (red line), $\tilde{H}_-$ (blue line)}\label{fig:01}
\end{figure}

\begin{figure}
\centering
\includegraphics[width=\columnwidth]{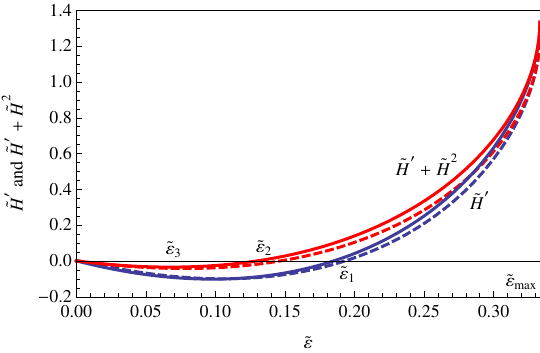}
\caption{Time derivative $\tilde{H}'$ and acceleration parameter $\tilde{H}' + \tilde{H}^2$ as functions of $\tilde{\veps}$ near limiting energy density} \label{fig:02}
\end{figure}

\begin{figure}
\centering
\includegraphics[width=\columnwidth]{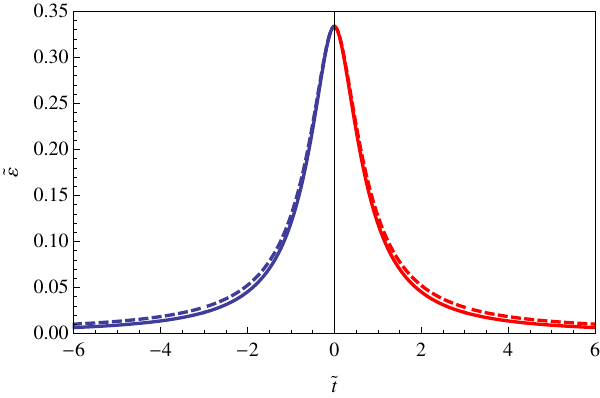}
\caption{Energy density $\tilde{\veps}$ as function of time $\tilde {t}$ near limiting energy density}
\label{fig:03}
\end{figure}

\begin{figure}
\centering
\includegraphics[width=\columnwidth]{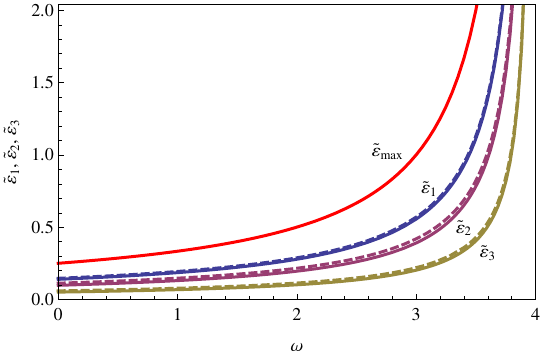}
\caption{Energies $\tilde{\veps}_{\rm max}$, $\tilde{\veps}_1$, $\tilde{\veps}_2$, $\tilde{\veps}_3$ as functions of parameter $\omega$ for flat and closed models}\label{fig:04}
\end{figure}

As follows from Fig.~2\footnote {The graphs in Fig.~2 are valid for both the
$\tilde{H}_-$-solution and the $\tilde{H}_+$-solution.} the derivative $ \tilde {H}'$
  decreases from its maximum value $4/3$ to zero at $\tilde{\veps}_1$. The acceleration parameter
  $\tilde {H}' + \tilde {H}^{2}$ is
  also reduced and vanishes at the energy density $\tilde{\veps}_2 = 0.130 = 0.39\, \tilde{\veps}_{\rm max} < \tilde{\veps}_1$. In the interval for the energy
  density ($\tilde{\veps}_2$, $\tilde{\veps}_{\rm max}$), the gravitational
  interaction has the character of repulsion, and at the density $\tilde{\veps}_2$ there is a transition from gravitational repulsion to attraction.
  With a further decrease in the energy density, the negative acceleration parameter reaches its minimum value $\tilde {H}' + \tilde {H}^{2}=-0.035$
  at $\tilde{\veps}_3 = 0.067 = 0.202\, \tilde{\veps}_{\rm max}$ corresponding to the maximum
  gravitational attraction force, which, as it decreases, approaches the gravitational attraction force of GR.
  The transition to the Friedmannian mode occurs when the value $\tilde{\veps}$ becomes much less than ${\tilde{\veps}}_{\rm max}$
  and the value $X$ is
  approaching 1; then at $\veps_{\rm vac}\ll \veps\ll \veps_{\rm max}$ according to (7) $\tilde{H} \sim \sqrt
{\tilde{\veps}}$, approximately such a transition occurs
  when $\tilde{\veps} =\tilde{\veps}_4 \sim 0.001\, \tilde{\veps}_{\rm max}$. By using the equation of energy conservation in dimensional
  form we obtain the dependence $\tilde{\veps} =\tilde{\veps}(\tilde {t})$ at extreme conditions presented in Fig.~3. Assuming
  that limiting energy density corresponds
   to $\tilde {t}=0$, we
   find an estimate for the moments of time $\tilde {t}_1=  \pm 0.642$, $\tilde {t}_2=  \pm 0.928$, $\tilde {t}_3=  \pm 1.541$,
   $\tilde {t}_4=  \pm 27.145$
   corresponding to $\tilde{\veps}_1$,
   $\tilde{\veps}_2$, $\tilde{\veps}_3$. $\tilde{\veps}_4$.
 By using obtained data we will estimate the time interval $\Delta {t}=
 ( \Delta {\tilde{t}}){ \sqrt{6f_0 \omega \alpha_G}}$ ($\Delta {\tilde {t}} = 2 {\tilde {t_4}}$) of transition
from the Friedmannian compression mode to the Friedmannian expansion mode. If we
assume that the magnitude of the limiting energy density is two orders of magnitude higher
than the density of a neutron star ($\alpha_G \sim {10^{-37}} ($kg$/$m~s$^2)^{-1}$),
$\omega =1$), we find by using the Stefan--Bolzmann law the limiting temperature $ {T_{\rm
max}} \sim 10^{13}$~K corresponding to the era of quark-gluon plasma.  We find in this
case  for the transition time from compression to expansion the following estimation $\Delta
{t}\approx 0.8 \cdot 10^{-3}$~s.

    As the numerical analysis for closed cosmological models ($k=+1$, $C_1=15$, $\omega =1$) shows, the
results for them at the beginning of the cosmological expansion are close to the
corresponding results for flat cosmological models. So the energy parameters have the
following values: $\tilde {\veps_1}=0,193=0,580 \tilde {\veps}_{\rm max}$,  $\tilde
{\veps_2}=0,147=0,441 \tilde {\veps}_{\rm max}$,  $\tilde {\veps_3}=0,977=0,232 \tilde
{\veps}_{\rm max}$.  The Hubble parameter vanishing at the limiting energy density
$\tilde{\veps}_{\rm max}$  reaches its maximum (minimum) value $\tilde{H}_{+}= 0.314$
($\tilde{H}_{-}= - 0.314$) at the energy density $\tilde{\veps}_1$.

    The behavior of cosmological solutions under extreme conditions near the limiting energy
density is dependent on the value of the parameter $\omega$, on which the value of the
limiting energy density $\tilde{\veps}_{\rm max}$  as well as the characteristic energies
$\tilde{\veps}_i$ ($ i=1,2,3$) depend. The dependence of the energy parameters on the
parameter $\omega$ in the case of flat ($k=0$) and closed ($k=+1$, $C_1=15$)
cosmological models is presented in Fig.~4.

  The analysis of open cosmological models near limiting energy density shows that their properties
under extreme conditions are close to them for cosmological models of flat and closed type,
fundamental differences occur in asymptotics like to GR. As an illustration, the solution for
the Hubble parameter in the case of an open model ($k=-1$, $C_1=15$, $\omega =1$) is
given in Fig.~5.

\begin{figure}
\centering
\includegraphics[width=\columnwidth]{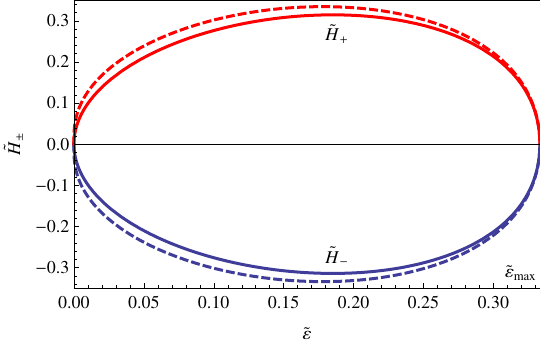}
\caption{Parameter $\tilde{H}=\tilde{H}_\pm$ as function of $\tilde{\veps}$ for $\omega=1$ and $k=0$ (solid lines) and for $\omega=1$, $k=-1$, and $C_1 = 15$  (dashed lines): $\tilde{H}_+$ (red line), $\tilde{H}_-$ (blue line)}\label{fig:05}
\end{figure}

      As the value of the limiting energy density ${\veps}_{\rm max}$ =  $(\omega \alpha_G)^{-1}\tilde{\veps}_{\rm max}$ increases (decreases), the value
of the limiting temperature also increases (decreases), which leads to corresponding changes
in the history of the early Universe dependent on two parameters: $\omega$ and $\alpha_G$.
In this case, the parameter $\alpha_G$ can play a role of fundamental physical constant in
evolution of the Universe. The value of parameter $\alpha_G$ is important also in
cosmological asymptotics defining the value of effective cosmological constant $\Lambda$
and as a result, a constant $b$. By using the accepted value of cosmological constant
$\Lambda=1.0905 \cdot 10^{-52}$ m$^{-2}$ and the following corresponding vacuum
energy density $\veps_{\rm vac}= 2\Lambda b \approx 2\Lambda f_0 = 5.25 \cdot
10^{-10}$ kg$/$m~s$^2$ we can obtain the following estimation $(1- \frac {b}{f_0}) =(4
\alpha_G \veps_{\rm vac})^{\frac{1}{2}}$.

\section{The gauge gravitation theory in the Riemann--Cartan space-time, gravitational interaction and torsion}

As follows from the isotropic cosmology studied above, the gravitational interaction within
the framework of GTRC under certain conditions may have the character of repulsion.  A
notable feature of the obtained cosmological solutions is the presence of an acceleration
stage at the beginning of cosmological expansion that can lead to corrections in the history of
early Universe. So in \cite{m10} the cosmological acceleration stage at the beginning of
cosmological expansion was introduced with the purpose to solve the problem of early
galaxy formation along with solving the Hubble tension problem, its appearance is based on
the hypothesis of early dark energy. Within the framework of the studied isotropic
cosmology built in the Riemann--Cartan space-time, dark energy does not exist, and
cosmological acceleration at the beginning of cosmological expansion is associated with the
existence of a limiting energy density for matter.

The physical consequences associated with gravitational interaction within
 the GTRC presented above are directly related to the role of space-time torsion generated
 by the energy-momentum tensor of gravitating matter (see e.g. \cite{k1, m9}).
If the limiting energy density exists in the nature, this should lead to important physical
consequences also in astrophysics. The properties of dense astrophysical objects with
energy densities comparable to the limiting energy density differ from what GR gives. The
fundamental consequence from a physical point of view is to prevent collapse and excludes
singular states with divergent energy density characteristic for black holes in GR. Significant
changes in the gravitational interaction in the case of astrophysical objects with energy
densities small compared to the limiting energy density take place when their spin moment
interacting with torsion is taken into account. It is noteworthy that the PGTG was created as
a generalization of GR to the case when the gravitational field has as sources, in addition to
the energy-momentum tensor, also the spin moment and the first, simplest PGTG was the
Einstein--Cartan theory, in the frame of which the torsion is connected directly by the spin
moment of matter. For definiteness, note that we consider the spin moment of astrophysical
objects as their tetrad spin moment, the most important contribution to which is the rotation
moment,
 and we stick to the covariant formulation of Hamilton's principle for oriented
particles and so-called spinning media in space-time with curvature and torsion developed in
\cite{a7, a8}. The point is that the interaction of spin moment of astrophysical objects with
space-time torsion leads to a modification of Newton's law of gravitational interaction. As
was shown within the framework of the so-called minimal GTRC \cite{m8}, which contains
the same uncertain parameters as isotropic cosmology, the interaction of vacuum torsion with
the rotational moments of astrophysical objects (stars, galaxies) leads to the appearance in
addition to the Newtonian gravitational attraction force, of an additional force caused by their
interaction \cite{m7}.  In addition to the vacuum torsion and the torsion generated by the
energy-momentum tensor, the torsion generated by own rotational moments of astrophysical
objects can play an important role in astrophysics. The search for GTRC that correctly
describes the interaction of torsion with the rotational moments of astrophysical objects is
very important for the theory of gravity.The physical consequences associated with the
interaction of torsion with the rotational moments of astrophysical objects may be of
fundamental importance in connection with the problem of dark matter.

\section{Conclusion}

The analysis of the solutions of isotropic cosmology carried out, constructed within the
framework of the gauge gravitation theory in Riemann-Cartan space-time, confirms the
conclusion that the gravitational interaction under certain conditions may have the character
of repulsion, this opens up possible ways to solve fundamental problems of the general
relativity theory. Being a direct development of GR, the GTRC is built on the basis of
generally accepted physical principles, including the principle of gauge invariance. The
fundamentally important place occupied in classical field theory, in theory of fundamental
physical interactions by the group of space-time translations and the Lorentz group, members
of the gauge group of GTRC determines the place of GTRC in the theory of gravitation. The
construction of a regular isotropic cosmology of accelerating Universe with limiting energy
density and limiting temperature makes it possible to exclude from the theory of gravitation
the concept of an initial cosmological singularity, which is unacceptable from a physical point
of view, as well as the concept of dark energy. Further studies of astrophysical objects that
have their own rotational moments within the framework of GTRC should give important
physical results for gravitation theory.

\subsection*{Data Availability Statement} The data that support the findings of this study
are openly available in \emph{Mendeley Data} at
https://doi.org/\hspace{0pt}10.17632/b2kvvxynvb, Ref.~\cite{Minkevich24Mendeley}.

\subsection*{Acknowledgments}

 The author is grateful to Dr. S. A. Vyrko for his help in carrying out this work.


\end{document}

%% file: preamble.tex
\usepackage{amsfonts,amssymb,cite}
\usepackage{graphicx}

\def\noi{\noindent}
\def\jnumber#1#2{\thispagestyle{empty} \noi\unitlength=1mm
    	\begin{picture}(178,10)
            \put(177,15){\llap{\large\it Preprint submitted to Grav. Cosmol.}}
                    \end{picture}}

\newcommand{\Title}[1]{\noi {{\Large\bf #1}}\\[1ex]}

\def\Aunames#1{\noi{\bf #1}}
\def\au#1{${}^{#1}$}
\def\Addresses#1{\medskip\noi \protect
	\begin{description}\itemsep -3pt {\it #1} \end{description}}
\def\adr#1#2{\item[${}^{#1}$]{\it #2}}

\newcommand{\Abstract}[1]{\vskip 2mm \begin{center}
        \parbox{16.4cm}{\small\noi #1} \end{center}\medskip}

\def\email#1#2{\footnotetext[#1]{e-mail: #2}\addtocounter{footnote}{1}}

\def\nqq{\hspace*{-2em}}

\def\cm{\hspace*{1cm}}

\usepackage{color}

\def\Jl#1#2{#1 {\bf #2},\ }

\def\ApJ#1 {\Jl{Astroph. J.}{#1}}
\def\CQG#1 {\Jl{Class. Quantum Grav.}{#1}}
\def\DAN#1 {\Jl{Dokl. AN SSSR}{#1}}
\def\GC#1 {\Jl{Grav. Cosmol.}{#1}}
\def\GRG#1 {\Jl{Gen. Rel. Grav.}{#1}}
\def\IJMPD#1 {\Jl{Int. J. Mod. Phys. D}{#1}}
\def\JETF#1 {\Jl{Zh. Eksp. Teor. Fiz.}{#1}}
\def\JETP#1 {\Jl{Sov. Phys. JETP}{#1}}
\def\JHEP#1 {\Jl{JHEP}{#1}}
\def\JMP#1 {\Jl{J. Math. Phys.}{#1}}
\def\NPB#1 {\Jl{Nucl. Phys. B}{#1}}
\def\NP#1 {\Jl{Nucl. Phys.}{#1}}
\def\PLA#1 {\Jl{Phys. Lett. A}{#1}}
\def\PLB#1 {\Jl{Phys. Lett. B}{#1}}
\def\PRD#1 {\Jl{Phys. Rev. D}{#1}}
\def\PRL#1 {\Jl{Phys. Rev. Lett.}{#1}}

\def\lal{&&\nqq {}}

\def\beq{\begin{equation}}
\def\eeq{\end{equation}}
\def\bear{\begin{eqnarray}}
\def\bearr{\begin{eqnarray} \lal}
\def\ear{\end{eqnarray}}
\def\earn{\nonumber \end{eqnarray}}

\def\yy{\\[5pt] {}}

